\documentclass[conference]{IEEEtran}
\IEEEoverridecommandlockouts
\usepackage{cite}
\usepackage{amsmath,amssymb,amsfonts}
\usepackage{algorithmic}
\usepackage{graphicx}
\usepackage{textcomp}
\usepackage{xcolor}

\usepackage{caption}
\usepackage{subcaption}

\def\BibTeX{{\rm B\kern-.05em{\sc i\kern-.025em b}\kern-.08em
    T\kern-.1667em\lower.7ex\hbox{E}\kern-.125emX}}
\begin{document}

\title{Exploiting Application-to-Architecture Dependencies for Designing Scalable OS
}

% \author{
% \IEEEauthorblockN{Yao Xiao, Nikos Kanakaris, }
% \IEEEauthorblockA{\textit{fff} \\
% \and
% \and
% \and
% \and
% \and

\author{
    \IEEEauthorblockN{Yao Xiao\IEEEauthorrefmark{1} \qquad Nikos Kanakaris\IEEEauthorrefmark{1}  \qquad Anzhe Cheng\IEEEauthorrefmark{1} \qquad Chenzhong Yin\IEEEauthorrefmark{1}\\ Nesreen K. Ahmed\IEEEauthorrefmark{2} \qquad Shahin Nazarian\IEEEauthorrefmark{1} \qquad Andrei Irimia\IEEEauthorrefmark{1} \qquad Paul Bogdan\IEEEauthorrefmark{1}}\\
    \IEEEauthorblockA{\IEEEauthorrefmark{1}University of Southern California, Los Angeles, CA, USA}
    \IEEEauthorblockA{\IEEEauthorrefmark{2}Cisco Research
}
}

\maketitle

\begin{abstract}
With the advent of hundreds of cores on a chip to accelerate applications, the operating system (OS) needs to exploit the existing parallelism provided by the underlying hardware resources to determine the right amount of processes to be mapped on the multi-core systems. However, the existing OS is not scalable and is oblivious to applications. We address these issues by adopting a multi-layer network representation of the dynamic application-to-OS-to-architecture dependencies, namely the NetworkedOS. We adopt a compile-time analysis and construct a network representing the dependencies between dynamic instructions translated from the applications and the kernel and services. We propose an overlapping partitioning scheme to detect the clusters or processes that can potentially run in parallel to be mapped onto cores while reducing the number of messages transferred. At run time, processes are mapped onto the multi-core systems, taking into consideration the process affinity. Our experimental results indicate that NetworkedOS achieves performance improvement as high as 7.11x compared to Linux running on a 128-core system and 2.01x to Barrelfish running on a 64-core system.
\end{abstract}


%\indent \textbf{Limitations of state-of-the-art approaches and motivations behind this work.} 
\section{Introduction}
The monolithic OS design is oblivious to applications and the computing platform, leading to performance issues: (1) It lacks scalability for multi-core systems~\cite{boyd2010analysis} due to cache coherence protocols, which synchronize threads in multi-threaded applications using locks to prevent inconsistent cache values~\cite{kadosh2024omparautomaticparallelizationaidriven, 10.1007/978-3-031-69577-3_9}. (2) Process communication in the kernel\footnote{We follow Linux's approach of treating threads as lightweight processes (LWPs), using 'processes' and 'threads' interchangeably.} is not suited for multi-core platforms~\cite{baumann2009your}. Network-on-Chip~\cite{marculescu2009chip, 8657366} is introduced to handle on-chip communications via message passing. (3) The general-purpose OS runs unnecessary processes, consuming time and resources. It offers diverse services, but only a small subset is relevant to most users.

% The current monolithic-based OS is oblivious to applications and the computing platform, which imposes performance issues:
% (1) The OS is not scalable to the multi-core systems~\cite{boyd2010analysis} due to cache coherent protocols. In multi-threaded applications, locks are commonly used to synchronize threads and prevent them from getting access to shared data structures. Cache coherent protocols ensure that only one cache is in exclusive mode to prevent multiple caches from having different values for one variable. 
% (2) The communication type of processes\footnote{In this paper, we borrow the idea from Linux to consider threads as light-weight processes (LWPs). Therefore, processes and threads are used interchangeably.} in a kernel does not match with the multi-core platforms~\cite{baumann2009your}. With the advent of multiple simple cores, it is found that the underlying communication hardware does not scale to multi-core systems. Therefore, Network-on-Chip \cite{marculescu2009chip} is proposed to handle on-chip communications via message passing. 
% (3) The current general-purpose OS runs unnecessary processes, consuming time and resources. While it offers diverse services for potential users, only a small subset is relevant to individual users.

%\indent \textbf{Challenges of the current approaches.} 
To address these issues, microkernels~\cite{liedtke1995micro, tanenbaum1987operating} were introduced. Unlike monolithic kernels, microkernels offer basic OS mechanisms like address space and thread management, as well as IPC. However, microkernel-based OSs face two key challenges:
% The rest of services like device drivers and file systems are provided in the user space as shown in Figure~\ref{fig:2}.  However, the two main challenges for microkernel OS are as follows: 
(1) User-level processes and servers exchange messages for IPC, which involve multiple round-trips. This can lead to slower responses, especially with a high message volume compared to monolithic OSs~\cite{9671685}. (2) Kernel memory block operations such as page zeroing and memory copies cost half the execution time of applications~\cite{calhoun2006optimizing, chen1994impact}. 

\begin{figure}[htpb]
\centering
\includegraphics[width=0.8\linewidth]{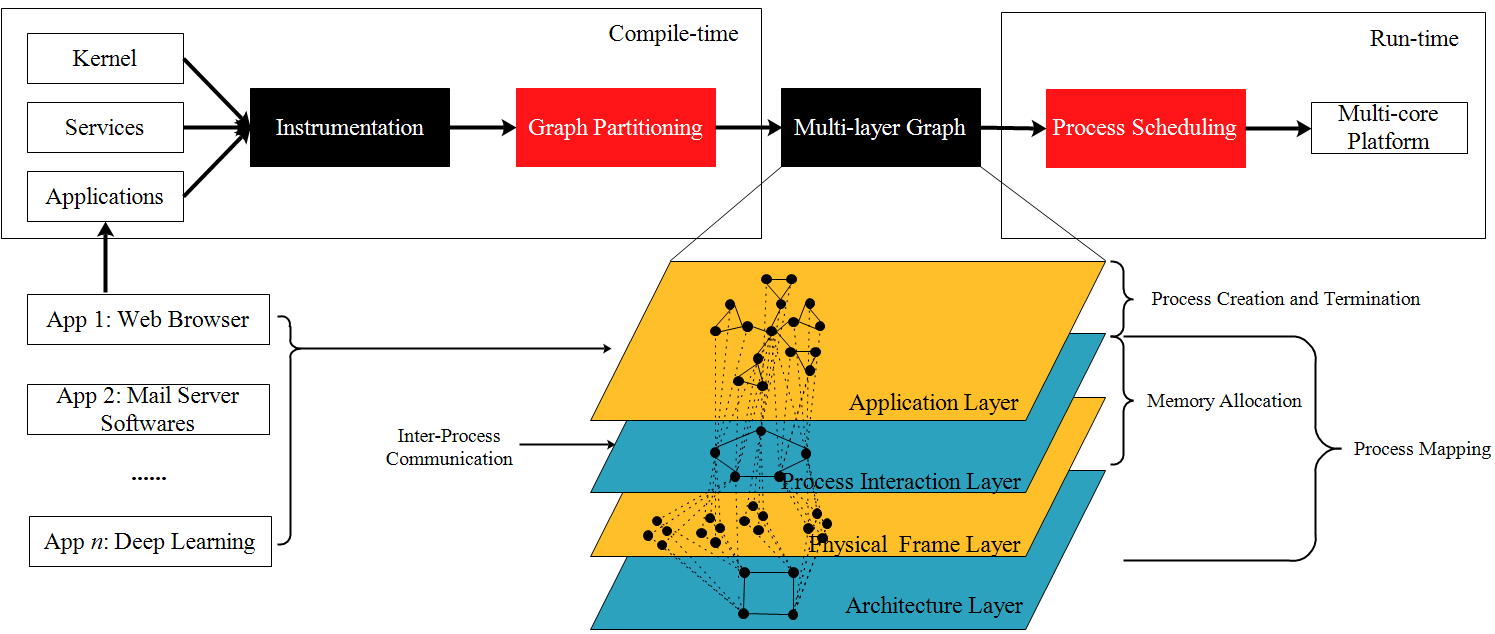}
\caption{\textbf{Overview of the NetworkedOS framework}. First at compile time, we build the multi-layer network to analyze the correlations among applications, kernel, and different services (e.g., device drivers and file systems). We then partition the application layer into optimal number of processes to be executed on the multicore platform. Next, based on the interactions among processes and the number of page frames utilized in each process, we propose a scheduling algorithm to map processes onto cores at run time.}
\label{fig:1}
\vskip -10pt
\end{figure}

%\indent \textbf{Key insights and contributions.} 
This paper aims to provide a design methodology for an efficient self-optimizing micro-kernel-based OS by introducing a novel multi-layer network abstraction that captures the dynamic interactions among applications and the OS. We achieve this by: (1) Running representative applications in the microkernel \textit{MINIX 3} to collect low-level traces; (2) Analyzing these traces for data and control dependencies, forming an interconnected multi-layer graph (MLG)~\cite{SIACHOS2025125290}; (3) Partitioning this MLG into dense clusters to represent distributed servers within the kernel; (4) Developing a scheduling algorithm that maps processes to cores based on their interaction patterns and memory affinity.
As shown in the multi-layer network of Figure~\ref{fig:1}, nodes in the application, interaction, physical and architecture layer represent dynamic instructions, processes running inside a kernel, physical frames translated from virtual memories and hardware cores, respectively~\cite{9397284}. Therefore, the multi-layer network of our NetworkedOS captures the following dynamic and heterogeneous dependencies: 
(1) \textbf{Process management} involves managing processes. Processes are sets of instructions executed by the OS.
(2) \textbf{Memory allocation} responsible for managing virtual memories for processes. Each process maintains a kernel stack to track virtual memory usage for stacks, heaps, and code/data segments.
(3) \textbf{Process mapping} assigns processes to hardware cores for execution. In our framework, this is represented by indirect edges connecting processes to cores through virtual memories.
(4) \textbf{Inter-process communication} handles shared data among processes.
% \indent With this framework, we try to answer the following questions in terms of applications to be executed and the underlying hardware architecture: (1) process creation: How to determine the number of processes with the help of applications and services (e.g., device drivers and file servers) to minimize the messages transferred via IPC; (2) process scheduling: How to schedule processes onto cores to exploit process affinity in order to reduce the amount of hops messages need to travel. \\ 

Towards this end, our main contributions are as follows: 
(i) The multi-layer network is aware of instructions running at the user level and encodes the correlations between the user level and kernel level;
(ii) We propose a novel graph partitioning algorithm to better create processes inside an OS to reduce the number of IPC;
(iii) We propose a process-to-cores scheduling algorithm that reduces the number of hops traveled by IPC messages and the number of memory block operations.

\vspace{-2mm}
\section{Methods}
\subsection{NetworkedOS}
\begin{figure}[htpb]
\centering
\includegraphics[width=0.8\linewidth]{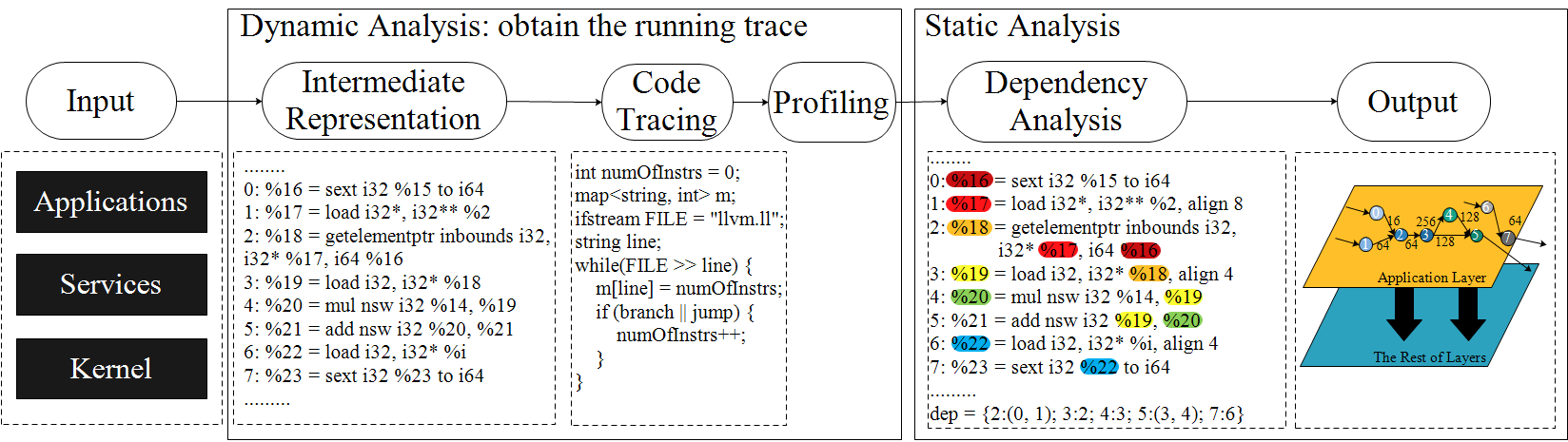}
%\vskip -5pt
\caption{\textbf{Overview of the multi-layer network construction}. We convert high-level languages into the corresponding dynamic low-level instructions. Using code tracing, analysis and profiling, we keep track of instructions in each basic block, analyze dependencies, and profile instructions to form an interconnected multi-layer network.}
\label{fig:3}
\vskip -10pt
\end{figure}

\noindent\textbf{Application Layer.}
We construct the application layer at the instruction level through a combination of \textbf{static} and \textbf{dynamic} analysis. This fine-grained method captures instruction inter-dependencies, enabling the formation of execution units or clusters. Dynamic instruction traces are generated, resolving branch instructions to track execution. After obtaining dynamic instructions, we use code tracing to identify instructions within basic blocks (Figure ~\ref{fig:3}). A container, \textit{map}, stores instruction keys within their basic blocks, with values representing instruction indices. We then profile the dynamic trace to measure memory retrieval and latency, using the \textit{rdtsc}() function for accuracy. Dependency analysis identifies register-dependent instructions. We maintain three containers: \textit{dest} and \textit{src} for destination and source registers by instruction index, and \textit{dep} for dependent instructions. The parser extracts source and destination registers and updates \textit{dep} when a source register matches a previous instruction's destination.
% We create the application layer at the instruction level using a combination of \textbf{static} and \textbf{dynamic} analysis. This fine-grained approach captures instruction inter-dependencies, enabling us to form multiple execution units or clusters. We dynamically accumulate instructions to generate accurate dynamic instruction traces, which include resolving branch instructions to track executed instructions.\\
% \indent After acquiring dynamic instructions, we use code tracing to identify instructions within basic blocks, as depicted in Figure ~\ref{fig:3}. We manage a container named \textit{map}, where keys represent specific instructions within their respective basic blocks, and values indicate the instruction index. Subsequently, we profile the dynamic trace to measure data retrieval from memory banks or cache levels and latency, utilizing the lightweight \textit{rdtsc}() function for accuracy.\\
% \indent Next, we conduct dependency analysis to identify register-dependent instructions. We maintain three containers: \textit{dest} and \textit{src} maps for tracking destination and source registers by instruction index, and a \textit{dep} container to map dependent instructions. As the parser processes an instruction, it extracts current source and destination registers. It checks if the source registers depend on destination registers from previous instructions, updating the \textit{dep} container to record dependencies when a match occurs.

\noindent\textbf{Process Interaction Layer.} Processes in the microkernel MINIX 3 are communicated via message passing. Therefore, in this layer, nodes and edges represent processes and direct communication, respectively. We monitor the kernel to keep track of the number of IPCs generated by applications, services, and the kernel. For example, in \textit{kernel/proc.c}, the \textit{mini\_send} function can be monitored to see a process (\textit{caller\_ptr}) sending messages (\textit{m\_ptr}) to another process (\textit{dst\_e}). In MINIX 3, IPC calls include \textit{send}, \textit{receive}, \textit{sendrec}, \textit{sendnb}, \textit{notify}, and \textit{senda}.
%In addition, although MINIX 3 has plenty of system calls, in essence, IPC using the function \textit{sendrec} is wrapped into these calls. \\

Regarding the dependencies between the application and the process interaction layer, inter-layer links exist between instructions and a process where they are executed.
%\indent As we will point out in Section IV, the number of processes and IPC calls may not be optimal in kernel to speed up applications. Therefore, we will reconstruct the inter-layer links by compile-time optimization of application execution time to map instructions to processes.

\noindent\textbf{Physical Frame Layer.}
 Physical frames mean physical page addresses in the memory subsystem. In this layer, nodes represent physical frames. Inter-layer links between the process interaction and physical frame layer represent whether a process can utilize the memory addresses. The MINIX 3 kernel maintains a function called \textit{pt\_writemap} whenever pages are allocated or freed, which records plenty of virtual memory regions for the address space. The virtual memory regions are then converted into physical frames for each process using page tables.  

\noindent\textbf{Architecture Layer.}
In the architecture layer, nodes represent hardware cores, and links represent communication hardware. In the kernel, we monitor the process-to-core mapping using physical frames on multi-core platforms. For instance, in \textit{kernel/proc.h}, \textit{p\_cpu} indicates the current process CPU, while \textit{p\_cpu\_mask} specifies allowed CPUs. In Section V, we highlight that MINIX 3's current multi-core process scheduling lacks considerations for process interactions, such as affinity and memory block operations. Therefore, we propose a new runtime scheduling algorithm to improve application speed.
\subsection{Compile-Time Optimization}
Forked processes in a multi-core application require efficient management and scheduling in the kernel to support inter-process communication (IPC). Deciding on the right number of processes is challenging: Too few result in limited parallelism and lower CPU utilization, while too many lead to excessive context switches and IPC overhead, affecting execution time. To address this complexity, we conduct compile-time analysis to partition the application layer (including applications, the kernel, and services) into clusters, determining the optimal number of processes. Here, `partition' refers to creating processes for instruction sequence mapping, and `clusters' represent the resulting process groups. We provide relevant definitions before explaining our optimization approach.\\
% \begin{figure}[t]
% \centering
% \includegraphics[width=0.5\textwidth]{figs/community.png}%, height=0.24\textwidth
% %\vskip -5pt
% \caption{\textbf{Instruction replication}. The nodes in the middle represent instructions whereas the nodes on the right represent processes. Enabling instruction replication to detect overlapping clusters can reduce the number of messages transferred.}
% \label{fig:4}
% \end{figure}
\indent \textbf{\textit{Definition 1}:} A multi-layer network is a weighted interconnected graph $\mathcal{G} = (V^i, E^{ij}, W^{ij} | i,j\in \{0, 1, 2, 3\})$ where $V^i$ is a set of vertices in the $i$th layer $ V^i = (v^i_k|k \in \{0,...,|V^i|\})$; $E^{ij}$ is a set of inter-links and intra-links between the $i$th and $j$th layers $E^{ij}=(e^{ij}_{kl}|k \in \{0,...,|V^i|\}, l \in \{0,...,|V^j|\})$; $W^{ij}$ is a set of weights for inter-links and intra-links between the $i$th and $j$th layers $W^{ij}=(w^{ij}_{kl}|k \in \{0,...,|V^i|\}, l \in \{0,...,|V^j|\})$.  \\
\indent \textbf{\textit{Definition 2}:} The quality function $T$, is defined to indicate whether a partitioning is superior according to the execution time of applications. \\
\indent \textbf{\textit{Definition 3}:} A partition state $s$ is represented as a mapping from nodes to clusters. Clusters contain a sequence of instructions that represent processes from a graph in the application layer.\\
\indent Therefore, using these definitions, we can formulate the compile-time optimization as follows:

% \begin{figure*}
% \centering
% \includegraphics[width=1\textwidth, height=0.25\textwidth]{figs/graph_2.png}
% %\vskip -5pt
% \caption{\textbf{Network partitioning}. Initially, all nodes in the network are placed into the same cluster. Next, each node in the network is placed either into a new cluster or the same cluster as its neighboring node. Once all of the nodes in the network are traversed, we merge nodes in the same cluster as one node and create a new network, which serves as the new network in the next pass.}
% \label{fig:5}
% \end{figure*}

\textbf{Given} a multi-layer network $\mathcal{G}$, \textbf{find} \textit{overlapping} clusters $N$, i.e., a mapping function $M:V^0\rightarrow V^1$, to \textbf{minimize} the quality function $T$
\vspace{-2mm}
\begin{equation}
\min\limits_{N, M}\hskip 8pt T(s) = W_{seq} + \frac{W_{par}}{N} + \frac{W_c}{N}
\end{equation}
\begin{equation}
W_{seq} = \sum_{\substack{
1\leq u\leq n\\
}} W_u, \neg \exists v \in \{1,..,s\}, v\neq u, d_u=d_v
\end{equation}
\begin{equation}
W_{par} = \sum_{\substack{
1\leq u\leq n\\
}} W_u, \exists v \in \{1,..,s\}, v\neq u, d_u=d_v
\end{equation}
\vspace{-5mm}
\begin{equation}
W_c = \sum_{\substack{
1\leq u\leq n\\
}} S_u
\end{equation}
\noindent where $N$ is the number of available cores; $W_u$ represents the sum of edge weights in the cluster $u$ ($W_u=\sum_{i\in u, j\in u} w^{00}_{ij}$); $d_u$ represents the depth of the cluster $u$ starting from the root; $W$ represents the total sum of edge weights in the application layer ($W=\sum_{ij}w^{00}_{ij}$); $S_u$ represents the sum of weights of edges adjacent to the cluster $u$. \\
%\indent talk about the meaning of each term and implications. draw a figure to explain each term and talk about why we need overlapping clusters.\\
\indent To enable high parallelism and reduce the amount of messages via IPC, overlapping clusters are detected, which allows instruction replication among processes. Overlapping clusters share some common instructions to ensure no synchronization of read-modify-write threads and messages transferred. \\
\indent In equations (1)-(4), $W_{seq}$ represents the amount of sequential work in the execution of applications, which is measured by the sum of edge weights in cluster $u$ if only one cluster exists at the depth of the cluster $d_u$; $W_{par}$ represents the parallel work in applications, which is measured by the amount of edge weights in parallel clusters with the same depth $d_u$; $W_c$ measures the amount of messages to be passed among processes. In essence, we extend the Amdahl's law in multi-core platforms \cite{hill2008amdahl} to accommodate our model. \\
\indent As we can see from the equation (1), two tradeoff scenarios exist: (i) between the number of processes and the amount of messages passing among processes. 
%If the number of clusters (processes) is equal to the number of nodes in the application layer, $W_c$ dominates. If the number of clusters is only one, then $f$ equals zero and $W_{seq}$ dominates. In either case, the application execution consumes a large amount of time. 
(ii) between the number of instructions replicated among processes and the amount of instructions in each parallel process. 
%If a large number of instructions are replicated among processes to reduce messages via IPC, then a single parallel process suffers as more work is required to be done. 
Hence, equation (1) decides an optimal way to partition the network and obtain the number of clusters. 

\indent We introduce the quality function $T$ to evaluate and optimize network partitions~\cite{duan2024structureawareframeworklearningdevice}. Given the computational challenges of exact optimization in large networks~\cite{duan2023leveragingreinforcementlearninglarge}, we propose a greedy algorithm to map instructions to processes and analyze process structures at multiple scales.
Starting from an initial state $s$ (where the whole network is a singleton), each node $n_i$ is randomly selected, and actions that maximize the function gain are taken, placing nodes in neighboring clusters or separate ones if the gain is positive. This process is repeated until no further gain ($\Delta T = T_{new} - T_{old}$) is possible.
Nodes within the same cluster are then merged, and the process repeats until a single node remains. This hierarchical approach yields partitions that minimize the quality function $T$.
% \begin{algorithm}
%     \caption{Greedy graph partitioning algorithm}
%   \begin{algorithmic}[1]
%     \STATE \textbf{INPUT}: An application-layer network
%     \STATE \textbf{OUTPUT}: A partition state $s$
%     \STATE Start with the network as one singleton cluster as an initial partition state $s$
%     \REPEAT
%         \STATE /* Function Optimization */
%         \FOR{each node $n_i$ in the input network}
%             \STATE Compute $T_{old}$
%             \STATE Place $n_i$ into a new cluster
%             \STATE Compute $T_{new}$ and $\Delta T = T_{new}-T_{old}$
%             \FOR{each neighboring node $n_j$ of $n_i$}
%                 \STATE Place $n_i$ into the same cluster as $n_j$
%                 \STATE Compute $T_{new}$ and $\Delta T = T_{new}-T_{old}$
%             \ENDFOR
%             \IF{One of $\Delta T$ is positive}
%                 \STATE Pick the largest one and update the network
%             \ELSE
%                 \STATE $n_i$ remains its cluster without modification
%             \ENDIF
%         \ENDFOR
%         \STATE /* Partition Aggregation */
        
%     \UNTIL{no function gains}
%   \end{algorithmic}
% \end{algorithm}
\vspace{-4mm}
\subsection{Run-Time Mapping}
% \begin{figure}
% \centering
% \includegraphics[width=0.5\textwidth, height=0.28\textwidth]{figs/mapping.png}
% \vskip -5pt
% \caption{\textbf{Process mapping of NetworkedOS}. Based on the three observations, greedy based run-time mapping takes into account the process affinity, core utilization, and kernel block operations.}
% \label{fig:6}
% \end{figure}

During runtime, mapping processes to multi-core platforms for application execution is crucial. Failing to consider process topology, physical memories, and underlying hardware in the kernel can lead to performance degradation for the following reasons: (1) Waiting for cache update; (2) Block memory operations between I/O devices and memory \cite{chen1994impact}; and (3) Optimize core utilization for multi-core efficiency.
% \begin{itemize}
%     \item Waiting for cache update: The multi-core platforms require the cache coherence protocol to have consistent data over private caches. A process later mapped to a different core may increase the time spent for the cache coherence protocol to fetch a cache line from the previous core. 
%     \item Block memory operations between I/O devices and memory \cite{chen1994impact}: As reported in \cite{chen1994impact}, in the microkernel, block memory operations in VM, IPC, and file systems constitute a large overhead for program execution time because a large portion of data is referenced and transferred between caches and the main memory.
%     \item Core utilization: In an extreme case, processes may be mapped only onto one core to exploit cache temporal and spatial locality. However, the rest of cores remain idle for a long time. Therefore, core utilization should be taken into account to fully exploit parallelism in multi-core systems.
% \end{itemize}

\indent To overcome the limitations, run-time mapping should exploit the parallelism in the process interaction layer and optimize for the job, while considering process interactions and resources. There are three observations to help us design a better run-time mapping algorithm: (1) Processes sharing data structures can map to one core to reduce cache coherence overhead and block operations; (2) Communicating processes can map to adjacent cores for shorter hops; (3) Independent processes can map to different regions of multi-core platforms to minimize message sharing paths.

\indent The run-time mapping algorithm takes the process interaction and physical frame layers from profiling and schedules process-to-core mappings to enhance application performance. The key criterion is minimizing time complexity, so we propose a greedy algorithm with a time complexity of $O(P)$, where $P$ is the number of schedulable processes. First, we check if a process is ready to schedule and track its communication with other processes. Based on observation 3, independent processes are mapped to distant cores, balancing loads on the multi-core platform’s communication substrate.
Next, we calculate execution times for processes sharing memory, comparing scenarios where they are mapped to the same or different cores. Mapping decisions follow observations 1 and 3. While mapping processes with shared memory to the same core reduces block operation overhead, excessive processes per core can hinder parallelism. We use equation (1) to estimate execution times for processes mapped to the same or different cores.

\vspace{-5mm}
\section{Evaluation}
%This section is organized as follows. In Section 6.1, we discuss the system setup and the operating systems with benchmarks we utilize to evaluate the validity of our framework. In Section VI.B, We illustrate the characteristics of our multi-layer model. In Sections VI.C and VI.D, we provide results for inter-process communication and system calls respectively. In Section VI.E, we evaluate our model on real HPC benchmarks to see the performance speedup. In Section VI.F, we analyze the scalability of our model with respect to the number of cores.

\noindent\textbf{System Setup.}
% \begin{figure*}[t]
% \centering
% \includegraphics[width=1\textwidth, height=0.23\textwidth]{figs/multi1.png}
% \vskip -5pt
% \caption{\textbf{Multi-layer network illustration for \textit{IS}}}
% \label{fig:7}
% \end{figure*}
% \begin{figure*}[t]
% \centering
% \includegraphics[width=0.9\textwidth, height=0.21\textwidth]{figs/multi2.png}
% \vskip -5pt
% \caption{\textbf{Multi-layer network illustration for \textit{BMI}}}
% \label{fig:8}
% \vskip -15pt
% \end{figure*}
We validated our results on a 64-processor system with 2-core 64-bit Intel Core$^{TM}$ i7-6600U processors (2.6 GHz), 16 GB RAM, and a 1 TB SSD. Each core has 128 KB L1 cache, 512 KB L2 cache, and 4 MB LLC, supporting 2 simultaneous threads per core. We adapted \textit{MINIX 3}'s microkernel to monitor application-kernel interactions, integrated them into our model and process mapping algorithm, and compared our model with three contemporary operating systems/kernels: unmodified \textit{MINIX 3} \cite{tanenbaum1987operating}, Linux kernel 4.18-rc4, and Barrelfish \cite{baumann2009multikernel}.
%We also adapted applications to ensure successful execution on Barrelfish while preserving core functionalities. \\

 The two HPC benchmarks we run are NAS Parallel Benchmarks 3.0 C code version and PARSEC 3.0 Benchmarks \cite{bienia2011benchmarking}. However, for NAS Benchmarks, the EP, BT, LU, MG, and SP benchmarks are excluded because Barrelfish cannot support features in these benchmarks such as thread local storage and file operations. Table 1 shows the benchmarks we used in this section.
%to measure the application performance and scalability in Sections VI.E and VI.F.
\begin{table}
\centering
\caption{Benchmarks}
\scalebox{1}{
\begin{tabular}{l|l} \hline\hline
Benchmark&Description \\ \hline
\textit{CG}&Conjugate gradient\\ \hline
%\textit{TRIAD}&Vector dot product operation&SHOC\cite{danalis2010scalable}\\ \hline
\textit{IS}&Integer sort\\ \hline
%\textit{SCAN}&Prefic sum algorithm&SHOC\cite{danalis2010scalable}\\ \hline
\textit{FT}&3D fast fourier transform\\ \hline
\textit{x264}&H.264 video encoding\\ \hline
\textit{fluidanimate}&Fluid dynamics for animation purposes\\ \hline
\textit{freqmine}&Frequent itemset mining\\ \hline
\textit{blackscholes}&Option pricing with PDE\\ \hline
\textit{streamcluste (sc)}&Online clustering of input streams\\ \hline
\textit{kms}&K-means clustering\\ \hline
\textit{PR}&PageRank algorithm to rank websites\\ \hline
\textit{BP}&Back-propagation in neural networks\\ \hline
\textit{CNN}&Convolutional neural networks\\ \hline
\textit{BMI}&Brain machine interface\\ \hline\hline
\end{tabular}
}
\end{table}
\vspace{-1mm}

\begin{figure}
     \centering
     \begin{subfigure}[]{\linewidth}
         \centering
         \includegraphics[width=\textwidth]{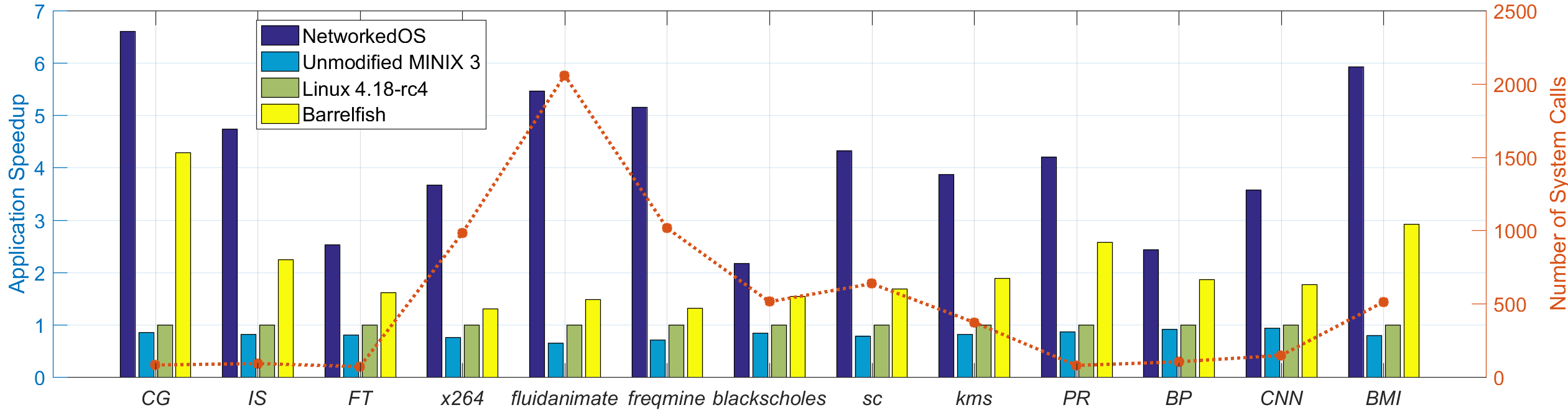}
     \end{subfigure}
     \hfill
     \begin{subfigure}[]{0.45\linewidth}
         \centering
         \includegraphics[width=\textwidth]{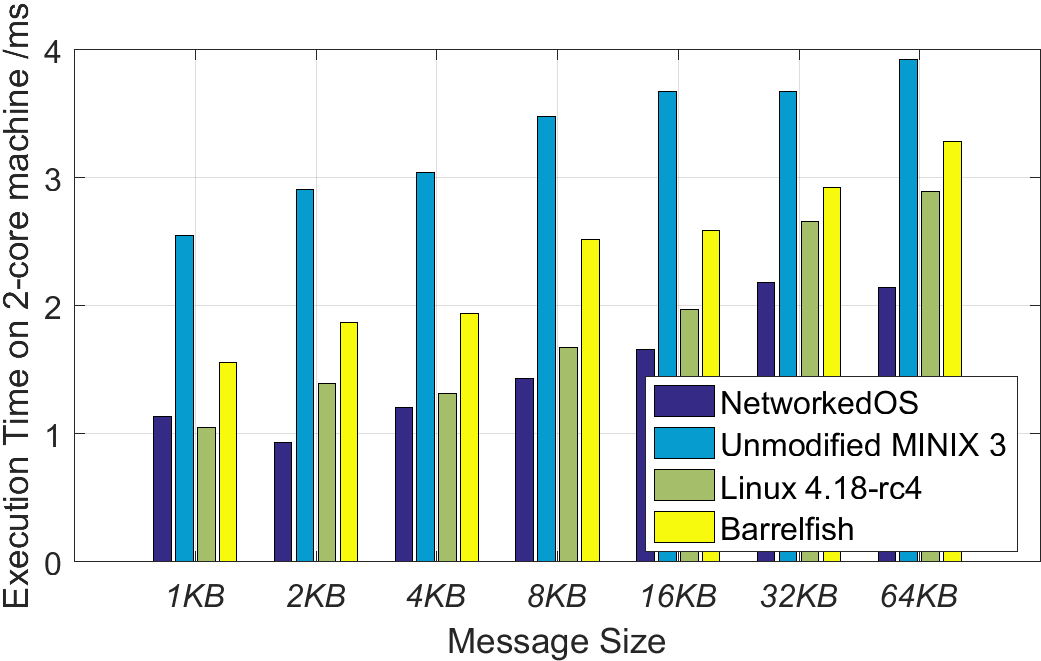}
         \label{fig:three sin x}
     \end{subfigure}
     \hfill
     \begin{subfigure}[]{0.5\linewidth}
         \centering
         \includegraphics[width=\textwidth]{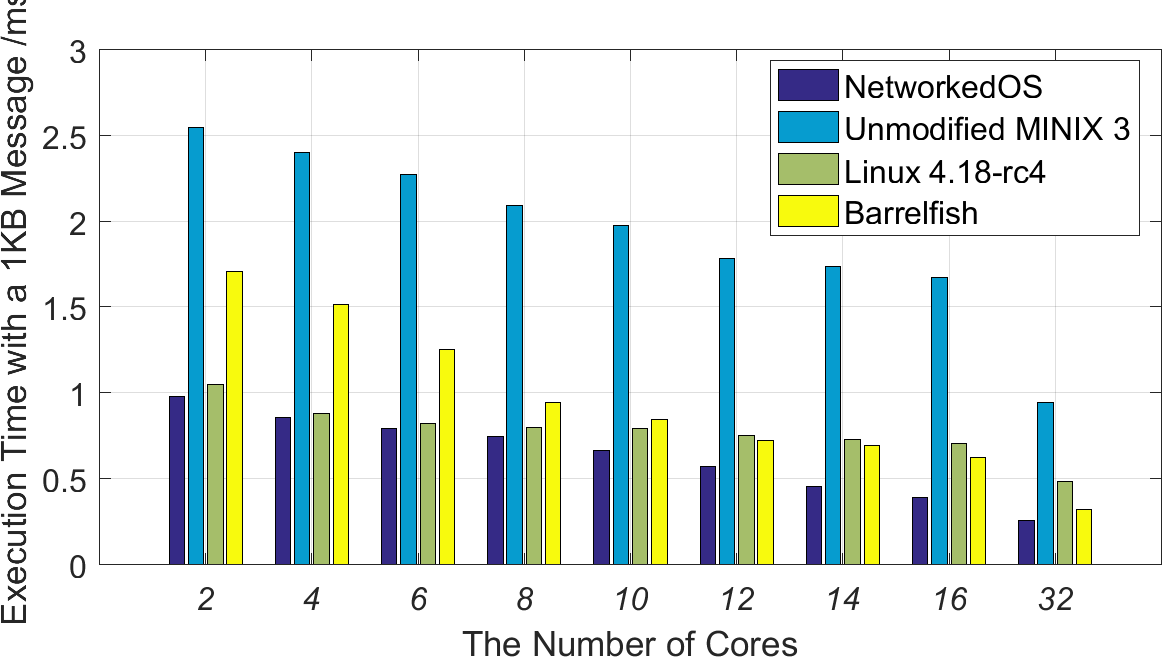}
         \label{fig:five over x}
     \end{subfigure}
    
\caption{\textbf{(Top)} Application speedup comparison. \textbf{(Bottom left)} Execution time on a 2-core machine. \textbf{(Bottom right)} Execution time with a 1KB message.}

\label{fig:9}
\label{fig:10}
\label{fig:11}
\end{figure}

\noindent\textbf{Inter-Process Communication.}
We run microbenchmarks to evaluate inter-process communication using \textit{ipc\_send} and \textit{ipc\_receive} in MINIX 3, \textit{pipe} in Linux, and \textit{interface} in Barrelfish via inter-dispatcher communication (IDC). The benchmarks fork processes and transfer messages of predefined lengths between pairs of processes. For example, with 4 processes, messages are transferred between processes 0 and 1, and between processes 2 and 3 simultaneously. 
We then vary the number of cores (2 to 32) and message sizes (1KB to 64KB). Figure~\ref{fig:10} (bottom left) shows execution times on a 2-core machine with message sizes ranging from 1KB to 64KB. Compared to the monolithic Linux kernel, the microkernel MINIX 3 incurs higher overhead due to increased message transfers between kernel and user mode when communicating with the file system. However, NetworkedOS reduces inter-process messages and execution time by grouping closely connected instructions into processes, resulting in up to a 3.4x reduction in execution time compared to unmodified MINIX 3 and a 1.35x reduction compared to Linux 4.18-rc4.

Figure~\ref{fig:11} (bottom right part) depicts execution times using fixed 1KB messages on a multi-core platform. Generally, Linux outperforms Barrelfish on systems with a limited number of cores, while Barrelfish excels on larger core counts, owing to its scalable multikernel design. Nevertheless, our approach intelligently maps processes to cores, making efficient use of multi-core platforms. Consequently, our approach achieves a 3.22x reduction in execution time compared to unmodified MINIX 3 and a 1.27x reduction compared to Barrelfish on a 32-core platform.

\noindent\textbf{Application Performance.}
Moreover, we conducted application-level benchmarks (see Table 1) on a 16-core platform, comparing performance against three baselines, with Linux 4.18-rc4 as the reference baseline. In Figure~\ref{fig:9} (top part), the red dotted line represents the number of executed system calls for each application, while the bar chart shows the corresponding application speedup.
For applications like \textit{FT}, \textit{PR}, \textit{BP}, and \textit{CNN}, characterized by an all-to-all communication pattern, our approach effectively reduces inter-process messages, even though the number of system calls is relatively low compared to PARSEC benchmarks.
In the case of \textit{CG} and \textit{IS}, our physical frame layer captures correlations between processes and memory addresses, optimizing core allocation for improved performance. 
For \textit{blackscholes}, Linux performs reasonably well due to low off-chip traffic and high parallelism, resulting in modest speedup improvements of only 1.47x for our approach and 1.14x for Barrelfish over Linux.
Regarding other PARSEC benchmarks with high system calls, cache misses, and off-chip traffic, our framework minimizes inter-process messages and leverages on-chip caches to map processes sharing the same memory allocations to the same core. Overall, our approach achieves up to 5.91x and 2.01x higher speedup compared to Linux and Barrelfish, respectively.
\begin{figure}
\centering
% \includegraphics[width=0.5\textwidth,height=0.25\textwidth]{figs/cg.png}
% \caption{\textbf{Application \textit{CG} Speedup}}
% \label{fig:13}

% \includegraphics[width=\linewidth]{figs/fluid.png}
% \caption{\textbf{Application \textit{BMI} Speedup}}
% \label{fig:14}

\end{figure}
%\vspace{-2mm}

% \noindent\textbf{Scalability.}
% Another important factor in OS is how well the OS scales over the number of cores. The baseline is Linux 4.18-rc4. While the compute-intensive benchmark \textit{CG} has few system calls and does not scale well, our approach could still recognize the underlying memory patterns and map processes onto cores to trade off between parallelism and the number of IPCs. 
% %As shown in Figure~\ref{fig:13}, NetworkedOS achieves 1.48x better speedup on a 64-core platform compared to Barrelfish. 
% % As shown in Figure~\ref{fig:14},
% NetworkedOS achieves 7.11x and 1.50x performance improvement over linux and Barrelfish, respectively, on a 128-core machine on the application \textit{BMI}.

% The application \textit{BMI} does not scale well in linux due to the thread communication overhead. However, NetworkedOS tries to minimize the communication cost, which, illustrated in Figure~\ref{fig:14}, achieves 7.11x and 1.50x performance improvement over linux and Barrelfish, respectively, on a 128-core machine.
%\vspace{-2mm}
\section{Conclusions}

%In this paper, we observe that the traditional monolithic kernel linux does not scale well on multi-core platforms whereas the microkernel such as MINIX 3 has serious performance issues due to numerous IPC messages between the user mode and the kernel mode. To mitigate the issue from the microkernel, we propose a multi-layer network theory in order to capture the underlying correlations between applications and the kernel. At the compile time, we first compile applications, the kernel, and services into LLVM IR instructions and construct the application layer by dependency analysis. We then monitor the activities from the kernel to construct the process interaction layer and physical frame layer. However, the number of processes forked by the kernel has the significant impact on performance. Therefore, based on the application layer, we propose an optimization model to decide the right amount of processes such that we fully exploit the hardware resources. At the run time, based on the memory locations referenced by processes and the interactions among processes, we design a greedy mapping algorithm from processes to cores in order to improve performance.\\
In this paper, we introduce NetworkedOS, a multi-layer network framework that captures correlations between applications and the kernel. We compile applications, the kernel, and services into LLVM IR instructions and build the application, process interaction, and physical frame layers through dependency analysis. Optimizing the number of forked kernel processes based on the application layer enhances hardware resource utilization. At runtime, we employ a greedy mapping algorithm to assign processes to cores for performance improvement based on memory locations and inter-process interactions. Our experiments validate our approach, showcasing improvements in inter-process communication, system calls, application performance, and scalability on a full system.

\noindent\textbf{Acknowledgement:} The authors acknowledge the support by the National Science Foundation (NSF) under the Career Award CPS-1453860, CCF-1837131, MCB-1936775, CNS-1932620 and award No. 2243104, Center for Complex Particle Systems (COMPASS), U.S. Army Research Office (ARO) under Grant No. W911NF-23-1-0111, DARPA Young Faculty Award and DARPA Director Award under Grant Number N66001-17-1-4044, an Intel Faculty Award and a Northrop Grumman grant.  The views, opinions, and/or findings in this article are those of the authors and should not be interpreted as official views or policies of the Department of Defense or the National Science Foundation.

\bibliographystyle{IEEEtran}
\bibliography{main}

\begin{thebibliography}{10}
\providecommand{\url}[1]{#1}
\csname url@samestyle\endcsname
\providecommand{\newblock}{\relax}
\providecommand{\bibinfo}[2]{#2}
\providecommand{\BIBentrySTDinterwordspacing}{\spaceskip=0pt\relax}
\providecommand{\BIBentryALTinterwordstretchfactor}{4}
\providecommand{\BIBentryALTinterwordspacing}{\spaceskip=\fontdimen2\font plus
\BIBentryALTinterwordstretchfactor\fontdimen3\font minus
  \fontdimen4\font\relax}
\providecommand{\BIBforeignlanguage}[2]{{%
\expandafter\ifx\csname l@#1\endcsname\relax
\typeout{** WARNING: IEEEtran.bst: No hyphenation pattern has been}%
\typeout{** loaded for the language `#1'. Using the pattern for}%
\typeout{** the default language instead.}%
\else
\language=\csname l@#1\endcsname
\fi
#2}}
\providecommand{\BIBdecl}{\relax}
\BIBdecl

\bibitem{boyd2010analysis}
S.~Boyd-Wickizer, A.~T. Clements, Y.~Mao, A.~Pesterev, M.~F. Kaashoek,
  R.~Morris, and N.~Zeldovich, ``An analysis of linux scalability to many
  cores.'' in \emph{OSDI}, 2010.

\bibitem{kadosh2024omparautomaticparallelizationaidriven}
\BIBentryALTinterwordspacing
T.~Kadosh, N.~Hasabnis, P.~Soundararajan, V.~A. Vo, M.~Capota, N.~Ahmed,
  Y.~Pinter, and G.~Oren, ``Ompar: Automatic parallelization with ai-driven
  source-to-source compilation,'' 2024. [Online]. Available:
  \url{https://arxiv.org/abs/2409.14771}
\BIBentrySTDinterwordspacing

\bibitem{10.1007/978-3-031-69577-3_9}
L.~Chen, A.~Bhattacharjee, N.~Ahmed, N.~Hasabnis, G.~Oren, V.~Vo, and
  A.~Jannesari, ``Ompgpt: A generative pre-trained transformer model
  for¬†openmp,'' in \emph{Euro-Par 2024: Parallel Processing}, J.~Carretero,
  S.~Shende, J.~Garcia-Blas, I.~Brandic, K.~Olcoz, and M.~Schreiber, Eds.\hskip
  1em plus 0.5em minus 0.4em\relax Cham: Springer Nature Switzerland, 2024, pp.
  121--134.

\bibitem{baumann2009your}
A.~Baumann, S.~Peter, A.~Sch{\"u}pbach, A.~Singhania, T.~Roscoe, P.~Barham, and
  R.~Isaacs, ``Your computer is already a distributed system. why isn't your
  os?'' in \emph{HotOS}, 2009.

\bibitem{marculescu2009chip}
R.~Marculescu and P.~Bogdan, ``The chip is the network: Toward a science of
  network-on-chip design,'' \emph{Foundations and Trends{\textregistered} in
  Electronic Design Automation}, 2009.

\bibitem{8657366}
Y.~Xiao, S.~Nazarian, and P.~Bogdan, ``Self-optimizing and self-programming
  computing systems: A combined compiler, complex networks, and machine
  learning approach,'' \emph{IEEE Transactions on Very Large Scale Integration
  (VLSI) Systems}, vol.~27, no.~6, pp. 1416--1427, 2019.

\bibitem{liedtke1995micro}
J.~Liedtke, \emph{On micro-kernel construction}, 1995, vol.~29, no.~5.

\bibitem{tanenbaum1987operating}
A.~S. Tanenbaum and A.~S. Woodhull, \emph{Operating systems: design and
  implementation}, 1987, vol.~2.

\bibitem{9671685}
G.~Ma, Y.~Xiao, M.~Capotă, T.~L. Willke, S.~Nazarian, P.~Bogdan, and N.~K.
  Ahmed, ``Learning code representations using multifractal-based graph
  networks,'' in \emph{2021 IEEE International Conference on Big Data (Big
  Data)}, 2021, pp. 1858--1866.

\bibitem{calhoun2006optimizing}
M.~Calhoun, S.~Rixner, and A.~L. Cox, ``Optimizing kernel block memory
  operations,'' in \emph{IEEE 4th Workshop on Memory Performance Issues}, 2006.

\bibitem{chen1994impact}
J.~B. Chen and B.~N. Bershad, ``The impact of operating system structure on
  memory system performance,'' in \emph{OSR}, vol.~27, no.~5, 1994, pp.
  120--133.

\bibitem{SIACHOS2025125290}
\BIBentryALTinterwordspacing
I.~Siachos, N.~Kanakaris, and N.~Karacapilidis, ``Software bug prediction using
  graph neural networks and graph-based text representations,'' \emph{Expert
  Systems with Applications}, vol. 259, p. 125290, 2025. [Online]. Available:
  \url{https://www.sciencedirect.com/science/article/pii/S0957417424021572}
\BIBentrySTDinterwordspacing

\bibitem{9397284}
Y.~Xiao, S.~Nazarian, and P.~Bogdan, ``Plasticity-on-chip design: Exploiting
  self-similarity for data communications,'' \emph{IEEE Transactions on
  Computers}, vol.~70, no.~6, pp. 950--962, 2021.

\bibitem{hill2008amdahl}
M.~D. Hill and M.~R. Marty, ``Amdahl's law in the multicore era,''
  \emph{Computer}, vol.~41, no.~7, pp. 33--38, 2008.

\bibitem{duan2024structureawareframeworklearningdevice}
\BIBentryALTinterwordspacing
S.~Duan, H.~Ping, N.~Kanakaris, X.~Xiao, P.~Zhang, P.~Kyriakis, N.~K. Ahmed,
  G.~Ma, M.~Capota, S.~Nazarian, T.~L. Willke, and P.~Bogdan, ``A
  structure-aware framework for learning device placements on computation
  graphs,'' 2024. [Online]. Available: \url{https://arxiv.org/abs/2405.14185}
\BIBentrySTDinterwordspacing

\bibitem{duan2023leveragingreinforcementlearninglarge}
\BIBentryALTinterwordspacing
S.~Duan, N.~Kanakaris, X.~Xiao, H.~Ping, C.~Zhou, N.~K. Ahmed, G.~Ma,
  M.~Capota, T.~L. Willke, S.~Nazarian, and P.~Bogdan, ``Leveraging
  reinforcement learning and large language models for code optimization,''
  2023. [Online]. Available: \url{https://arxiv.org/abs/2312.05657}
\BIBentrySTDinterwordspacing

\bibitem{baumann2009multikernel}
A.~Baumann, P.~Barham, P.-E. Dagand, T.~Harris, R.~Isaacs, S.~Peter, T.~Roscoe,
  A.~Sch{\"u}pbach, and A.~Singhania, ``The multikernel: a new os architecture
  for scalable multicore systems,'' in \emph{SOSP}, 2009, pp. 29--44.

\bibitem{bienia2011benchmarking}
C.~Bienia and K.~Li, \emph{Benchmarking modern multiprocessors}.\hskip 1em plus
  0.5em minus 0.4em\relax Princeton University Princeton, 2011.

\end{thebibliography}

\end{document}